\documentclass[aps,pra,twocolumn,10pt,nofootinbib,superscriptaddress]{revtex4-2}
\usepackage{amsmath,amssymb,amsthm,mathtools,bm}
\usepackage[hidelinks]{hyperref}
\usepackage{physics}
\usepackage{microtype}
\usepackage[T1]{fontenc}
\usepackage{lmodern}

\usepackage{booktabs} 
\usepackage{tikz} 

\newcommand{\Prb}{p}

\usepackage{xcolor}

\usepackage{multirow}
\usepackage{csquotes}
\usepackage{graphicx}
\usepackage{algorithm}
\usepackage{algorithmic}

\newtheorem{theorem}{Theorem}
\newtheorem*{theorem*}{Theorem}
\usepackage[capitalise]{cleveref}

\usepackage[caption=false]{subfig}

\begin{document}
\title{Quantum oracles give an advantage for identifying classical counterfactuals}

\author{Ciarán M.~Gilligan-Lee}
\affiliation{Department of Physics \& Astronomy, University College London, Gower Street, London, WC1E 6BT, UK}
\affiliation{Spotify, Dublin, Ireland}

\author{Y{\`i}l{\`e} Y{\=\i}ng}
\affiliation{Perimeter Institute for Theoretical Physics, Waterloo, Ontario, Canada, N2L 2Y5}
\affiliation{Department of Physics and Astronomy, University of Waterloo, Waterloo, Ontario, Canada, N2L 3G1}

\author{Jonathan Richens}
\affiliation{Google DeepMind, London, UK}

\author{David Schmid}
\affiliation{Perimeter Institute for Theoretical Physics, Waterloo, Ontario, Canada, N2L 2Y5}

\begin{abstract}
We show that quantum oracles provide an advantage over classical oracles for answering classical counterfactual questions in causal models,
or equivalently, for identifying unknown causal parameters such as distributions over functional dependences. 
In structural causal models with discrete classical variables, observational data and even ideal interventions generally fail to answer all counterfactual questions, since different causal parameters can reproduce the same observational and interventional data while disagreeing on counterfactuals. Using a simple binary example, we demonstrate that if the classical variables of interest are encoded in quantum systems and the causal dependence among them is encoded in a quantum oracle, coherently querying the oracle enables the identification of all causal parameters---hence all classical counterfactuals. We generalize this to arbitrary finite cardinalities and prove that coherent probing 1) allows the identification of all two-way joint counterfactuals \(\Prb(Y_x=y, Y_{x'}=y')\), which is not possible with any number of queries to a classical oracle, and 2) provides tighter bounds on higher-order multi-way counterfactuals than with a classical oracle. This work can also be viewed as an extension to traditional quantum oracle problems such as Deutsch--Jozsa to identifying more causal parameters beyond just, e.g., whether a function is constant or balanced. Finally, we raise the question of whether this quantum advantage relies on uniquely non-classical features like contextuality. We provide some evidence against this by showing that in the binary case, oracles in some classically-explainable theories like Spekkens' toy theory also give rise to a counterfactual identifiability advantage over strictly classical oracles. 
\end{abstract}

\maketitle

\section{Introduction}

The notion of a counterfactual---a hypothetical alternative to an event that did not in fact occur---sits at the core of modern causal reasoning \cite{pearl2009causality, vlontzos2023estimating}. Counterfactual questions underpin scientific explanation \cite{mucesh2024nature} and decision-making \cite{gilligan2020causing}, and are increasingly central in applications such as personalized medicine \cite{richens2020improving, reynaud2022d}, policy evaluation \cite{zeitler2023non, o2025spillover, o2025local, van2023estimating, andreu2024contrastive}, and algorithmic fairness \cite{kusner2017counterfactual}. In all of these settings, counterfactuals provide a language for attribution (``Was $X$ a cause of $Y$ in this particular case?''), for comparing alternative actions (``Would a different treatment have led to a better outcome?''), and for defining normative desiderata (``Would this decision have been the same if the individual’s sensitive attribute were changed?''). 

Recent developments in formal theories of causality, most notably the structural causal models framework pioneered by Judea Pearl \cite{Pearl2009}, have provided a precise mathematical language for studying counterfactuals. Central to Pearl's formalism is the notion of \emph{identifiability}, i.e., whether certain causal questions can be answered---or identified---from certain types of data. One foundational result \cite{bareinboim20201on} is the existence of counterfactual questions, such as ``given that a patient has certain symptoms, would they not have developed them had a certain disease been treated?'', that cannot be identified from data collected by passive observation or even active intervention. 

Here, we explore classical counterfactuals from a quantum perspective and ask whether those that cannot be identified using classical resources become identifiable when given access to quantum resources.  By ``classical counterfactuals'' (henceforth simply ``counterfactuals'' for short), we mean the kind of counterfactuals one can define in classical causal models. We do not consider more general questions regarding quantum counterfactuals such as the ones explored in Ref.~\cite{banerjee2025counterfactualquantummeasurements,suresh2024semanticscounterfactualsquantumcausal}.   

 In Pearl's framework, the causal influence of one variable on another is expressed mathematically by a function, and one's knowledge about this causal influence is described by a probability distribution over the possible functions.\footnote{See also Ref.~\cite{von2021algorithmic,padh2023stochastic, balke1994counterfactual,ansanelli2025resourcetheorycausalinfluence} for the importance of considering probability distributions over functions (rather than the stochastic maps they define) in contexts other than counterfactuals.} By shifting from the language of causal models to distributions over functions, we frame our question as an oracle problem that generalizes traditional oracle problems such as Deutsch--Jozsa~\cite{DeutschJozsa} 
in the sense that instead of having an oracle with a fixed function that one tries to identify, our oracle is sampled from a distribution that we try to identify characteristics of.

In this framing, we answer our question in the affirmative by showing that coherently probing a quantum oracle allows for the identification of certain aspects of the distribution over these functions---which correspond to counterfactuals---that cannot be identified classically. 
After formalizing this quantum advantage, we also derive its limits, showing for instance that there exist counterfactuals that cannot be uniquely identified with arbitrary independent queries to a classical or quantum oracle. Nevertheless, we further show that quantum oracles provide tighter bounds on higher-order multi-way counterfactuals than those achievable with a classical oracle.

Finally, we investigate whether this advantage is intrinsically quantum by exploring if it relies on uniquely non-classical features, like contextuality, or if it only relies on weakly non-classical features, like coherence. We provide evidence for the latter by showing that in dimension 2, one can also find oracle advantages in theories (like Spekkens' toy theory \cite{spekkens2007evidence,epistricted}) that are not strictly classical, but that are consistent with the notion of classical-explainability known as generalized noncontextuality~\cite{spekkens2005contextuality,Schmid2024structuretheorem}. This is analogous to what has been shown~\cite{johansson2017efficient} for other oracle-based advantages like Deutsch--Jozsa's. This raises interesting questions about how to compare resources (like oracles) across theories, and how our generalized oracle framing could be used to investigate whether strongly non-classical features, like contextuality, can be tied to quantum computational advantages.

\section{Non-identifiability of counterfactuals}\label{identifiability}

\begin{figure}[htb!] 
\centering
        \includegraphics[width=0.3\linewidth]{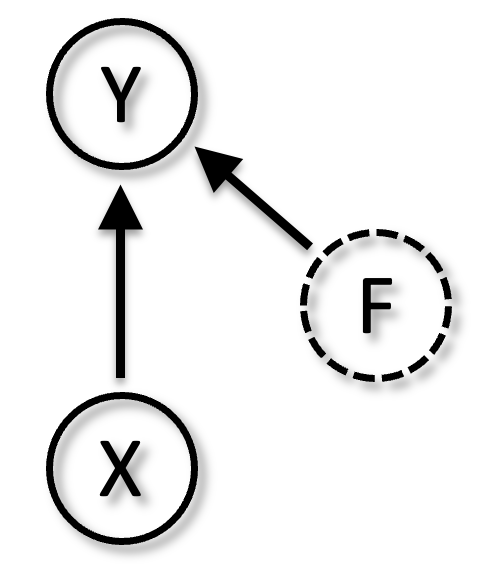}
        \caption{The DAG we consider, where $F$ controls the dependence of $Y$ on $X$. $X$ and $Y$ are observed, while $F$ is not.}
        \label{fig:DAG}
\end{figure}

Here we show by an explicit example that in Pearl's classical causal modeling framework, observational and interventional data do not generally allow for identification of counterfactual distributions, as multiple parametrizations of the causal phenomenon can be consistent with the observed data. Consider the simple case of a causal structure represented by the directed acyclic graph (or DAG) in Fig.~\ref{fig:DAG}, with binary observed variables $X,Y$, 
and a latent variable $F$ distributed according to $\Prb(F)$.  $F$ can be thought of as determining the functional dependence of $Y$ on $X$.\footnote{Formally, $F$ is a response-function variable~\cite{balke1994counterfactual,von2021algorithmic,padh2023stochastic}. Moreover, all results in this paper can be extended to the case where $X$ and $Y$ have an unobserved common cause. See \cref{app:cc} for details.}  There is no loss of generality in taking a four-valued $F$ as there are only four possible deterministic functions from a binary $X$ to binary $Y$: identity (denoted $\mathbb I$), flip (denoted $\mathbb{F}$), reset to zero (denoted $\mathbb R_0$), and reset to one (denoted $\mathbb R_1$).

How one answers counterfactual questions in each given run relies on one's knowledge $\Prb(F)$ (in that run) of what function governs the dependence of $Y$ on $X$. Consequently, we wish to estimate this distribution. We imagine that one does so using both observational and interventional data on $X$ and $Y$. For example, consider the counterfactual question $\Prb(Y_{X=1}=0 \mid Y=0, X=0)$, meaning ``given $X$ was observed to be 0 and $Y$ was observed to be 0, what is the probability that $Y$ would still be 0 had I intervened to set $X=1$?''  It can be computed via
\begin{align}\label{eq:counterfactual}
\Prb(Y_{X=1}=0 \mid Y=0, X=0) =\frac{\Prb(F={\mathbb R_0})}{\Prb(Y=0 \mid X=0)}
\end{align}
which follows by noting that $Y=0$ and $X=0$ implies that either $F$ is $\mathbb I$ or $\mathbb R_0$, and within this context the only way $Y=0$ when $X=1$ is if $F={\mathbb R_0}$. (See \cref{app:cc} for a brief introduction to counterfactual estimation in Pearl's framework.)

Since the marginal $\Prb(X)$ does not contain information about $\Prb(F)$, the observational data that is useful for finding $\Prb(F)$ is captured by conditional distributions, namely $\Prb(Y=0|X=0)$ and $\Prb(Y=0|X=1)$ (since $\Prb(Y=1|X=0)$ and $\Prb(Y=1|X=1)$ are fixed by normalization, $\sum_y \Prb(Y=y|X=x)=1$. Furthermore, the interventional data\footnote{The interventional conditional (or do-conditional) $\Prb(Y|{\rm do}(X))$ denotes the probability that $Y$ takes a certain value when the value of $X$ is set by the intervention instead of its causal parent.} on $X$ gives no new information, as $\Prb(Y|{\rm do}(X))=\Prb(Y|X)$ here. Thus, 
the only (independent) constraints one learns about $\Prb(F)$ are
\begin{subequations}\label{eq:obseg1}
\begin{align} 
    \Prb(Y=0|X=0) = \Prb(F={\mathbb I}) + \Prb(F={\mathbb R_0}), \\
    \Prb(Y=0|X=1) = \Prb(F={\mathbb F}) + \Prb(F={\mathbb R_0}).
\end{align}
\end{subequations}
But $\Prb(F)$ contains three free parameters---four, minus normalization, $\sum_f \Prb(F=f)=1$. Thus, there can exist different distributions over $F$ with the same conditional (and hence interventional) distributions, but different counterfactuals, even in this simple scenario. Since we do not have enough constraints to solve for $\Prb(F)$, we cannot necessarily answer counterfactual questions, such as the one in Eq.~\eqref{eq:counterfactual}. 
For example, the equal mixture of $\mathbb I$ and $\mathbb F$ gives the same values of $\Prb(Y=0|X=0)$ and $\Prb(Y=0|X=1)$ as the equal mixture of $\mathbb R_0$ and $\mathbb R_1$, and the former assigns 0 while the latter assigns 1 to the counterfactual $\Prb(Y_{X=1}=0 \mid Y=0, X=0)$.
Hence, there are counterfactuals that are not uniquely \emph{identifiable} from observational and interventional data without further constraints. 

If one can identify the causal model parameter 
$\Prb(F)$, then one can identify all counterfactuals. Consequently, the arguments in this paper can equivalently be viewed as being about causal parameter estimation
rather than full counterfactual estimation. 

\section{Identifying classical counterfactuals with quantum resources} \label{quantum identifiability}

\begin{figure}[htb!]
    \centering
    \includegraphics[width=0.45\textwidth]{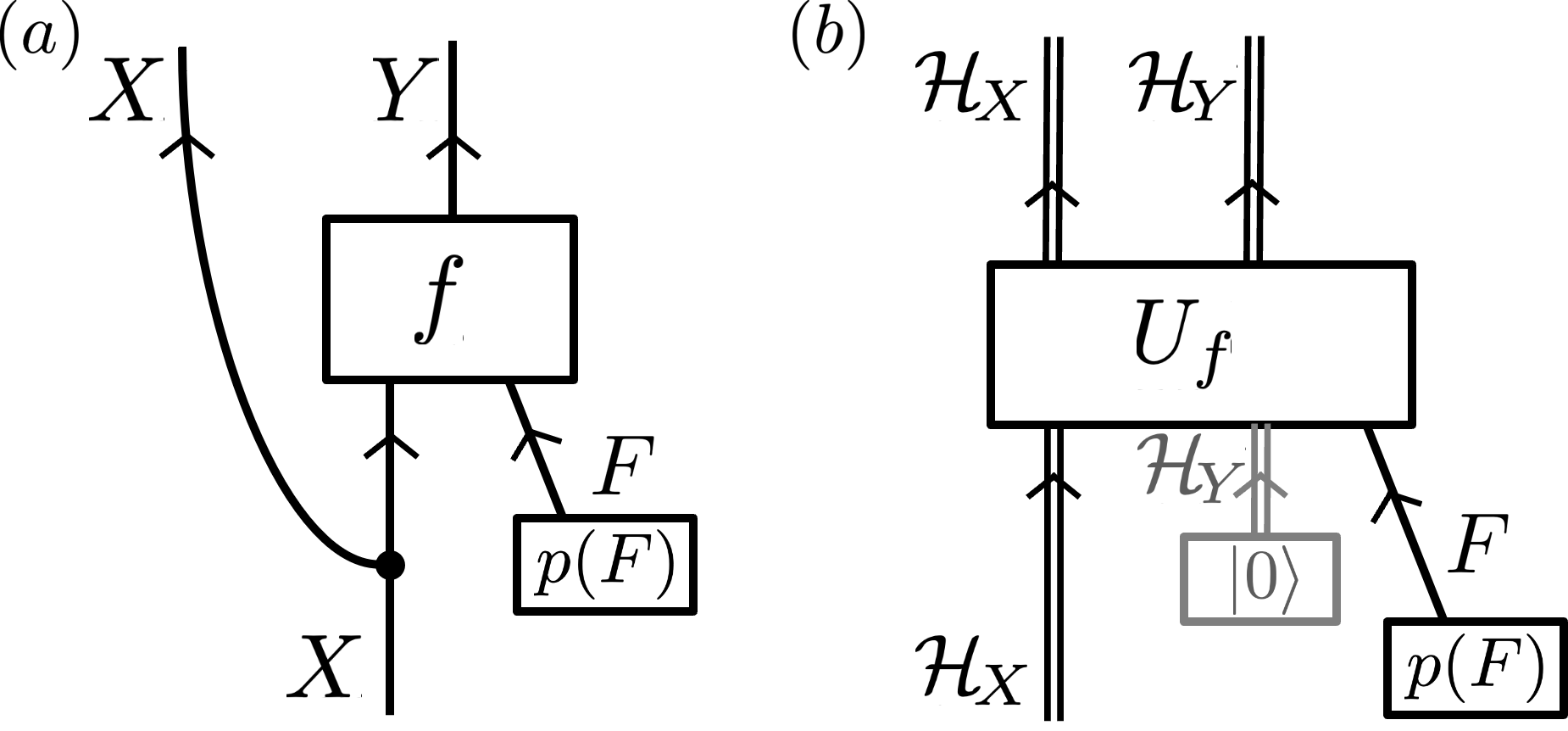}
    \caption{a) Classical scenario, where $F$ determines the functional dependence of $Y$ on $X$. b) Analogous quantum scenario, where $F$ determines the unitary dependence of $Y$ on $X$.
    }
    \label{fig:dilation}
\end{figure}

The DAG in \cref{fig:DAG} can equivalently be expressed as a string diagram, as shown \cref{fig:dilation}(a). There, we imagine the $X$ variable is copied (denoted by the black dot) to be an output of the string diagram to emphasize that it is an observed variable. 

Imagine now that the values of $X$ and $Y$ are instead encoded in quantum systems. That is, the classical variable $X$ is associated with a Hilbert space ${\cal H}_X$ equipped with a preferred basis (which we take to be the computational basis) whose basis states $\ket{0},\ket{1},...$ represent the possible values of $X$, and similarly for $Y$. Here, as $X$ and $Y$ are binary, perfect encoding can be achieved with ${\cal H}_X$ and ${\cal H}_Y$ being qubits.  Moreover imagine now that the causal dependence of $Y$ on $X$ is governed by a unitary process rather than a classical functional dependence. Analogous to before, the specific unitary is drawn from a set of four possibilities, depending on the value of a latent classical variable $F$. We can view this unitary as a \emph{quantum oracle} that one queries to estimate $\Prb(F)$---or equivalently, to try to identify the counterfactuals computed using $\Prb(F)$. The quantum oracle acts in the computational basis as
\begin{equation} \label{unitary}
U_f:\ \ket{x}_X\ket{0}_Y \mapsto \ket{x}_X\ket{f(x)}_Y.
\end{equation}
A unitary circuit depicting this process is shown in Fig.~\ref{fig:dilation}(b). 
It is a deterministic process, and the only place uncertainty enters is through $\Prb(F)$.  Also note that although we cannot simply copy a general input on ${\cal H}_X$ (due to the no-cloning theorem), the action of the oracle ensures that the value of the classical variable $X$ encoded in ${\cal H}_X$ does get copied in the computational basis.

In this way, we can see the problem of counterfactual identification we encountered earlier as an oracle problem. 
Unlike traditional quantum oracle problems where there is a fixed $f$ for every query of the oracle and the aim is to learn about properties of the fixed function $f$, here, 
the particular function used in each query is sampled according to the distribution $\Prb(F)$ and the aim is to learn about this distribution. (Nevertheless, we discuss in Section~\ref{sec:oracle} how the traditional quantum oracle problems can be seen as a special case of the one here.)

Analogously, we can view the classical scenario before as using  a classical oracle $C_f: x \mapsto \big(x,f(x)\big)$. We have seen previously that it is not possible to find $\Prb(F)$ (or equivalently to identify all counterfactuals) with any number of queries to this classical oracle.

In contrast, we now show that from observational and interventional data from queries to the quantum oracle, 
we can identify the entire distribution $\Prb(F)$, and hence can identify all counterfactuals.

First, consider the intervention preparing ${\cal H}_X$ 
in a computational basis state and consider measuring the output on ${\cal H}_Y$ in the computational basis, for which
\begin{subequations} \label{eq:0001}
   \begin{align}
   \label{eq:10q} \Prb\big(\bra{0}_Y \mid \, {\rm do} (\ket{0}_{\!X}\!)\big) = \Prb(F={\mathbb I})+\Prb(F={\mathbb R_0}),\\
      \label{eq:01q} \Prb\big(\bra{0}_Y \mid \, {\rm do} (\ket{1}_{\!X}\!)\big) = \Prb(F={\mathbb F})+\Prb(F={\mathbb R_0}).
\end{align} 
\end{subequations}
These two constraints are equivalent to  \cref{eq:obseg1} in the case with classical oracles. 

Now, however, we can obtain a third constraint on $\Prb(F)$, linearly independent of the two above.  Specifically, consider the intervention on ${\cal H}_X$ that prepares it in the $\ket{+}$ state. Then, the probability that a measurement on the joint system ${\cal H}_X\otimes{\cal H}_Y$ in the Bell basis yields the $\bra{\Phi^+}=\frac{1}{\sqrt{2}}(\bra{00}+\bra{11})$ outcome is:
\begin{align}
&\Prb\Big(\bra{\Phi^+}_{XY} \Big|\, {\rm do} (\ket{+}_{\!X}\!) \Big) \nonumber \\
= \, &
\Prb(F={\mathbb I})+\frac{1}{4}\Prb(F={\mathbb R_0})+\frac{1}{4}\Prb(F={\mathbb R_1}).
\end{align} 
Together with the normalization constraint that $\sum_f \Prb(F=f)=1$, this allows us to exactly compute $\Prb(F)$, which allows us to answer any counterfactual questions, including $\Prb(Y_{X=1}=0 \mid Y=0, X=0)$, which we could not do previously. 

 (Note that with the classical oracle we could also make a joint measurement on both variables $X$ and $Y$ after the intervention and the application of the oracle, and compute the distribution over the outcomes of the joint measurement conditioned on the intervention, but this does not provide any new information relative to measuring only $Y$, since the value of $X$ will always be the same as the initial value prepared for $X$ by the intervention.) 

Thus, we see that quantum oracles give an advantage over strictly classical oracles for identifying counterfactuals (or, equivalently, causal parameters like $\Prb(F)$).

\section{Higher cardinalities} \label{section: higher order}
Moving beyond binary variables, let $X,Y\in\{0,1,\dots,n-1\}$. We assume without loss of generality that $X$ and $Y$ have the same cardinality, since cases with different cardinalities can always be embedded into ones where both sets are assigned the larger of the two cardinalities. Then, it is sufficient for the cardinality of the response-function variable $F$ to be $n^n$ since there are $n^n$ different functions from $X$ to $Y$.  

Again, the conditionals under an intervention on $X$ are the same as the observational conditionals classically, which are
$
\Prb(Y=y\mid X=x)=
\sum_{f}\Prb(F) \,\delta_{y,f(x)}.
$
Following Pearl~\cite{pearl2009causality}, when there is no confounding, this is equal to the counterfactual distribution $\Prb(Y_{x}=y)$, and thus, a one-way counterfactual distribution can always be identified via observations or interventions in our causal structure.

A conditioned counterfactual like the one we saw before is obtained by 
\begin{align}
  \Prb(Y_{x'}{=}y'\mid X{=}x,Y{=}y) =
  & \frac{\sum_{f}\Prb(F) \,\delta_{y,f(x)} \delta_{y',f(x')}}{\sum_{f}\Prb(F) \,\delta_{y,f(x)}}
\end{align}
Here, the numerator is equal to the joint distribution over two counterfactual events~\cite{BareinboimETAL}, also called cross-world counterfactuals~\cite{pearl2016Causal,Richardson2013SingleWI}, namely $\Prb\big(Y_{x}{=}y,\ Y_{x'}{=}y'\big)$. We refer to such joint distributions as two-way joint counterfactuals for short. The above conditional counterfactual can now be expressed as
\begin{align}
    \Prb(Y_{x'}{=}y'\mid X{=}x,Y{=}y) =
  & \frac{\Prb\big(Y_{x}{=}y,\ Y_{x'}{=}y'\big)}{\Prb(Y=y\mid X=x)}.
\end{align}
The two-way joint counterfactuals contain more information about $\Prb(F)$ than the one-way counterfactuals $\Prb(Y_{x}=y)$. While the one-way counterfactuals are given by observational and interventional data directly,  with a classical oracle $f(x)$, for any $n\geq 2$, the two-way joint counterfactual distributions $\Prb\big(Y_{x}{=}y,\ Y_{x'}{=}y'\big)$ cannot always be identified, since different distributions can give rise to the same observational and interventional data. This can be seen by realizing that our $n=2$ example can be generalized to $n>2$, as shown in \cref{cident}.

\begin{theorem}\label{thm:twoway}
With quantum oracles as defined in \cref{unitary}, one can always identify two-way joint counterfactuals,
$\Prb\big(Y_{x}{=}y,\ Y_{x'}{=}y'\big)$, and so also the conditional counterfactuals $\Prb(Y_{x'}{=}y'\mid X{=}x,Y{=}y)$, for any cardinality of $X$ and $Y$, which is not always possible with classical oracles.
\end{theorem}
\noindent We prove this in \cref{qident}.

There is, however, a limit to this quantum advantage for counterfactual identifiability when we consider three-or-more-way joint counterfactuals (which are even more informative about $\Prb(F)$ than the two-way ones) such as $\Prb\big( Y_{x}{=}y,\ Y_{x'}{=}y', \ Y_{x''}{=}y'' \big)= \sum_{f}\Prb(F) \,\delta_{y,f(x)} \delta_{y',f(x')} \delta_{y'',f(x'')}$

\begin{theorem}\label{thm:limit}
$n$-way joint counterfactuals may not be fully identified when $n>2$ even with an arbitrary number of (independent) queries to the quantum oracle. 
\end{theorem} 
\noindent We prove this in \cref{qcident}. (For $n=2$, one can always identify any counterfactual with quantum oracles since, as shown earlier, one can learn $\Prb(F)$ completely.)

Note that we consider only independent queries to the oracle; we leave it an open question whether there are still limits to such identifiability problems if one has access to many copies of an oracle and can probe them in parallel (perhaps even feeding in entangled states as inputs). 

In classical causal models, when full identifiability is not possible, researchers have explored partial identification---upper and lower bounds on the counterfactual distribution of interest \cite{pearl2009causality, vlontzos2023estimating, padh2023stochastic}. While Theorem~\ref{thm:limit} shows there are limits to the counterfactuals that can be uniquely identified using quantum oracles, we prove in Appendix~\ref{app:partial} that they can provide tighter bounds on $n$-way joint counterfactuals than the known partial identifiability bounds attainable with classical oracles. 

\begin{theorem} \label{thm:n-way}
For $n := \max\{|\mathcal X|,|\mathcal Y|\}\geq2$, there exist $n$-way joint counterfactuals that can be more tightly bounded using quantum oracles than using classical oracles.
\end{theorem}

   There is thus a general advantage of quantum oracles over classical oracles in partial identification for $n\geq 2$. We leave for future work the question of quantifying how much tighter the bounds using quantum oracles is compared to classical oracles in identifying  counterfactuals.

\section{Extending traditional quantum oracle problems} \label{sec:oracle}

Traditional oracle problems, where $f$ is a fixed function, can in fact be viewed as a special case of identifying causal parameters, namely $\Prb(F)$, and thus as a special case of our problem. Specifically, $\Prb(F)$ is there promised to be a special kind of distribution, namely a deterministic (point) distribution. That is, $\Prb(F)$ is equal to 1 for a specific $f$ and 0 for all others, and one is trying to learn more about this distribution. In the case where one aims to find the exact function $f$ such that $\Prb(F=f)=1$, this is equivalent to completely identifying $\Prb(F)$, while the case where one is merely trying to learn certain properties of $f$ is equivalent to learning a coarse-grained feature of $\Prb(F)$, such as the two-way or three-way joint counterfactuals discussed earlier in Section~\ref{section: higher order}. 

For example, in Deutsch--Jozsa's algorithm~\cite{DeutschJozsa}, under the further promise that $\Prb(F)$ is supported either entirely on constant functions or entirely on balanced functions, one tries to determine a coarse-grained feature of $\Prb(F)$, namely whether $\sum_{f\in \mathrm{cons}}\Prb(F=f)$ (or equivalently $\sum_{f\in \mathrm{bal}}\Prb(F=f)$) is equal to 1, where `cons' denotes the set of constant functions and `bal' denotes the set of balanced functions. Similarly, in Simon's algorithm~\cite{Simon1997}, under the further promise that $\Prb(F)$ is supported only on functions $f$ for which there exists a nonzero bit string $s$ such that $f(x)=f(y)$ iff $y=x\oplus s$, the coarse-grained feature of $\Prb(F)$ one tries to identify is the value of this hidden period $s$, namely the unique $s$ such that $\sum_{f\in\mathcal F_s}\Prb(F=f)=1$, where $\mathcal F_s$ denotes the set of functions with period $s$. In contrast, in Bernstein--Vazirani's algorithm~\cite{BernsteinVazirani}, one tries to completely identify $\Prb(F)$. Here, $\Prb(F)$ is further promised to satisfy $\Prb(F=f)=0$ unless $f(x)=x\cdot s$ for some bit string $s$ (where $x$ and $s$ are bit strings and $\cdot$ denotes the inner product), and one then seeks to determine which such $f$ (or equivalently, which value of $s$) satisfies $\Prb(F=f)=1$.

Recently, there has been work on oracle problems where the oracle is sampled from a given distribution~\cite{rosmanis2023quantumsearchnoisyoracle,Shukla_2023}. There, $\Prb(F)$ is promised to be non-zero for a particular class of functions, and also to have a particular form, and one then aims to learn more about $\Prb(F)$. Again, our problem can be seen as a generalization of this in the sense that we allow arbitrary functions and $\Prb(F)$.\footnote{ Oracles sampled from a given distribution have also been studied in cryptography research (e.g., Ref.~\cite{Boneh2011,YamakawaZhandry2021}) and in complexity research (e.g., Ref.~\cite{Aaronson2007,agarwal2025cautionarynotequantumoracles} ). However, the connections between our work and those works are less clear.}

Thus, these considerations may be of independent interest for linking quantum computing and causal inference. 

\section{Spekkens Toy Theory and the quantum nature of the advantage}
We proved a quantum-over-classical advantage for identifying certain counterfactuals. However, this advantage relies on the assumption that the classical and quantum systems being compared have the same dimension, and on the idea that systems are `classical' iff they are described by classical probability theory---or more formally, by the simplicial generalized probabilistic theory~\cite{lee2015computation}.
There are other theories that can sensibly be said to be {\em classically-explainable}~\cite{SchmidGPT,Schmid2024structuretheorem}, such as those consistent with the principle of generalized noncontextuality~\cite{spekkens2005contextuality,schmid2020unscrambling}, and in these theories, the analogue of a $d$-dimensional quantum system generally involves a higher-dimensional classical variable. For example, in an epistemically restricted theory~\cite{spekkens2007evidence,epistricted,bartlett2012reconstruction,Catani2023whyinterference} such as Spekkens' toy theory~\cite{spekkens2007evidence}, the analogue of a qubit is given by two classical bits. If one considers oracles defined within this theory, one has perfect identifiability in the case where $n=2$. 
One can see this immediately by the fact that all of the quantum processes in our binary example are reproduced by Spekkens' toy theory. 

This is very similar to the situation with oracle-based advantages for quantum computing, as discussed in Ref.~\cite{johansson2017efficient}. There too, although quantum oracles give an advantage over strictly classical oracles for Deutsch--Jozsa's and Simon's algorithms, the authors show that if one uses a toy-bit type of oracle which is somewhere between a usual classical oracle (which computes a classical function) and a quantum oracle (which allows the classical functions to be queried coherently), then one finds no quantum advantage. 
The key lesson is that care is needed when comparing resources across quantum and classical theories, and that quantum theory does not provide an advantage for Deutsch--Jozsa's and Simon's algorithms relative to all classically-explainable theories.

Here,  at least in the binary case,  whether or not one attributes the advantage we proved for counterfactual identifiability (or causal parameter estimation) to be an instance of genuine nonclassicality depends on whether or not one compares to strictly classical theories, or to the broader set of classically-explainable theories such as the set of noncontextual theories. 

 When $n>2$, the quantum oracle $U_f$ can (for some $f$) lie outside the stabilizer subtheory of quantum theory, and so it is not yet clear whether our argument extends beyond the $n=2$ case. (Recall that in prime dimensions, every stabilizer subtheory is a noncontextual theory~\cite{catani2017spekkens,Schmid2022Stabilizer}.) 

Regardless, it is interesting to delineate the scope of identifiable counterfactuals in various theories, be they classical, quantum, or foil theories. Our work thus raises the question of whether quantum theory allows for {\em genuinely nonclassical} advantages for causal inference---that is, advantages over all classically-explainable theories.

\section*{Acknowledgments}

We thank Robert W. Spekkens for pointing out that the advantage in the binary case exists also in his toy theory. We thank Yujie Zhang for discussions on quantum oracles and the stabilizer subtheory of quantum theory. DS and YY were supported by Perimeter Institute for Theoretical Physics. Research at Perimeter Institute is supported in part by the Government of Canada through the Department of Innovation, Science and Economic Development and by the Province of Ontario through the Ministry of Colleges and Universities. 
YY is also supported by the Natural Sciences and Engineering Research Council of Canada (Grant No. RGPIN-2024-04419).

\bibliography{bibliography}

\appendix

\section{Response function formulation of structural causal models, confounding, and counterfactual estimation} \label{app:cc}

In the main text, we have followed the response function formulation of causal models \cite{balke1994counterfactual,von2021algorithmic,padh2023stochastic}. We now show how this formulation is formally related to Pearl's structural causal model formulation \cite{pearl2009causality}.

Let $X\in[n]:=\{0,1,\dots,n-1\}$ and $Y\in[m]$ be discrete variables. 
If $X\!\to Y$, in Pearl's structural causal model framework \cite{pearl2009causality} we can write $Y=f(X,W), X=g(V)$, where $W,V$ are exogenous noise terms distributed according to $\Prb(W), \Prb(V)$, and $f,g$ are deterministic functions. Any unobserved confounding between $X$ and $Y$ is manifested as statistical associations between $W$ and $V$. 

As $X,Y$ are discrete, it is shown in Chapter $8$ of \cite{pearl2009causality} that we can reformulate this into an equivalent setting where the $W,V$ are replaced by discrete response variables, one each for $X$ and $Y$: $(R_X,R_Y)$ (see recent work on this in \cite{padh2023stochastic, von2021algorithmic}). Intuitively, there are a finite number of distinct functions that map discrete $X$ to discrete $Y$, and we can think of the exogenous $W$ as (randomly) determining which function is applied. We can thus partition each $W$ into finitely many regions corresponding to these functions, and define a new discrete random variable $R$ that indicates which region $W$ falls into.

In the response function setting~\cite{von2021algorithmic}, a (possibly confounded) causal model for $X\!\to Y$ is specified by exogenous variables $(R_X,R_Y)$ and a family of deterministic response functions for $Y$:
\begin{equation}
Y \;=\; f_{R_Y}(X),\qquad f_{r}\in [m]^{[n]}.
\end{equation}
The population is determined by a mixture over functions with weights $\Prb(R_Y=f)$ such that $\sum_f\Prb(R_Y=f)=1$.  That is, the variable $F$ in the main text essentially plays the role of $R_Y$ here.

If there is unobserved confounding, then $R_X$ and $R_Y$ are statistically dependent, otherwise they are independent 
$\Prb(X{=}x,Y{=}y) \;=\; 
\sum_{r_x,f} \Prb(R_X{=}r_x,R_Y{=}f)\; \delta_{x=r_x}\delta_{y=f(x)}.$

Even when there is unobserved confounding (a latent common cause), and hence statistical dependence between the response variables for $X$ and $Y$, the results in this paper still go through, as we allow interventions on $X$, which break the association between response variables, and bring us back to a situation that is mathematically equivalent to one without unobserved confounding.  That is, to transport our results in the main text to the case where there is unobserved confounding, we only need to replace observational conditionals with do-conditionals. 

\subsection{Counterfactual estimation}\label{app:counter}
In Chapter $7$ of \cite{pearl2009causality}, Pearl provides a three-step procedure for estimating counterfactuals $\Prb(Y_x | \text{evidence})$: 
\begin{enumerate}
\item \textbf{Abduction}: first, update knowledge about $\cal F$, denoting the set of all exogenous variables in the structural causal model, given the observed evidence, 
\item \textbf{Action}: intervene on $X$ to set it to $x$, 
\item \textbf{Prediction}: use the posterior over latent variables $\Prb({\cal{F}} \mid \text{evidence})$ inferred in the abduction step, together with the intervened model in which $X$ is set to $x$, to compute the outcomes of $Y$ that would result under this intervention. The counterfactual distribution $\Prb(Y_x \mid \text{evidence})$ is then obtained by averaging the resulting $Y$ values over the posterior distribution of $F$.

\end{enumerate}

To see how this three-step process allows for the estimation of counterfactual distributions, recall our example from Section~\ref{identifiability}, where we want to estimate $\Prb(Y_{X=1}=0 \mid Y=0, X=0)$. In this case the evidence is $Y=0,X=0$, and we need to update our knowledge of $\Prb(F)$ under this to get the posterior $\Prb(F \mid Y=0,X=0)$. To do this, note that $Y=0$ and $X=0$ implies that either $F$ is $\mathbb I$ or $\mathbb R_0$. Hence $\Prb(F \mid Y=0,X=0)$ assigns zero probability to $F= \mathbb R_1$ and $F= \mathbb F$, while $\Prb(F=\mathbb I \mid Y=0, X=0) = \frac{\Prb(F=\mathbb I)}{\Prb(F=\mathbb I)+\Prb(F=\mathbb R_0)}$ and $\Prb(F=\mathbb R_0 \mid Y=0, X=0) = \frac{\Prb(F=\mathbb R_0)}{\Prb(F=\mathbb I)+\Prb(F=\mathbb R_0)}$. Next, we intervene on $X$ to set it to $1$. Finally, we sample from $\Prb(F \mid Y=0,X=0)$ and combine with $X=1$ to compute the probability that $Y=0$. Here, the only way $Y=0$ when $X=1$ under $\Prb(F \mid Y=0,X=0)$ is if $F={\mathbb R_0}$, which occurs with probability $\Prb(\mathbb R_0 \mid Y=0, X=0) = \frac{\Prb(F=\mathbb R_0)}{\Prb(F=\mathbb I)+\Prb(F=\mathbb R_0)}.$

\section{Non-identifiability of two-way joint counterfactuals with classical oracles} \label{cident}

To demonstrate the non-identifiability, all one needs to do is provide two causal models that agree on observations and interventions but disagree on two-way joint counterfactuals. Consider one model where the distribution of functions is the equal mixture of all the discard-the-input and prepare-a-fixed-output maps, and another model where the distribution is the equal mixture of all permutation maps. In both models, one has $\Prb(Y=y|X=x)=1/n$ for any $y$ and any $x$, since in the first case the output $y$ is randomly sampled independent of $x$, and since in the second case a random permutation is equally likely to take $x$ to any possible $y$.
So both models agree on observations and interventions. But in the second model, when $x\neq x'$ the two-way joint counterfactuals satisfy $\Prb(Y_x=y, Y_{x'}=y)=0$, while this is non-zero in the first model. Hence both models agree on interventions and observations, but disagree on certain joint counterfactual distributions---showing such distributions cannot in general be identified using classical oracles.

\section{Proving Theorem~\ref{thm:twoway}: identifying two-way joint counterfactuals with quantum oracles}\label{qident}

We use the same set-up as the one in \cref{fig:dilation}.

 Consider a general pure state $\ket{\psi}=\sum_x \alpha_x \ket{x}$ as the state on ${\cal H}_X$ prepared by the intervention. For a fixed $f$, the joint state on ${\cal H}_X\otimes {\cal H}_Y$ after the corresponding unitary $U_f$ is $\sum_x \alpha_x \ket{x}\ket{f(x)}$.
Then, for a given distribution over $f$, the joint state after the transformation is 
\begin{align}
\rho_{XY} = \sum_{x,x'} \alpha_x \alpha_{x'}^* \sum_{f} \Prb(F=f)\ketbra{x}{x'} \otimes \ketbra{f(x)}{f(x')}  
\label{eq:master}
\end{align}
The matrix elements of $\rho_{XY}$ in the computational basis are then
\begin{align}
  \bra{x}\langle y|\rho_{XY}|y'\rangle \ket{x'}
  &= \alpha_x\alpha_{x'}^*
     \sum_f \Prb(F=f)\delta_{f(x)=y, f(x')=y'} \nonumber\\
  &= \alpha_x\alpha_{x'}^*\,\Prb\big(f(x){=}y,\ f(x'){=}y'\big)
\end{align}

Then, if we chose the initial state $\ket{\psi}$ such that $\alpha_x\alpha_{x'}^*\neq 0$, by performing tomography on ${\cal H}_X\otimes {\cal H}_Y$, we can identify all two-way joint counterfactuals by 
\begin{equation}
  \label{eq:C-from-Lambda}
 \Prb\big(f(x){=}y,\ f(x'){=}y'\big) = \frac{1}{\alpha_x\alpha_{x'}^*}
  \bra{x}\langle y|\,\rho_{XY}\,|y'\rangle\ket{x'}.
\end{equation}

\section{Proving Theorem~\ref{thm:limit}: non-identifiability of three-way joint counterfactuals with quantum or classical oracles} 
\label{qcident}

First, we give an explicit example where two causal models, or more specifically, two distributions over functions $\Prb(F)$, can give different three-way joint counterfactuals while agreeing on observational/interventional data and two-way joint counterfactuals when $n>2$.

In this example, $n=3$. Model A: \(\Prb(F)\) is the uniform distribution over all deterministic functions
\(f:\{0,1,2\}\to\{0,1,2\}\). Model B: \(\Prb(F)\) is the uniform distribution over the \(9\) affine-linear
functions of the form
\(
f_{U,S}(x)=U+Sx \pmod 3,
\)
where \(U,S\in\{0,1,2\}\).
Both models give $\Prb(Y_x=y)=1/3$ for any $x$ and $y$, and $\Prb(Y_x=y, Y_{x'}=y')=1/9$ for any $x\neq x'$ and any $y$,$y'$, but differ on the three-way joint counterfactuals:
\begin{align}
\Prb(Y_0{=}0,Y_1{=}1,Y_2{=}2)=
\begin{cases}
1/27,&\text{Model A},\\
1/9,&\text{Model B}.
\end{cases} 
\end{align}

With the classical oracle, the three-way joint counterfactuals are not identifiable in this example because one only has access to $\Prb(Y,X)$ (from which $\Prb(Y|X)=\Prb(Y|{\rm do} (X)$ can be derived).

With a quantum oracle, from \cref{eq:master}, we know that the counterfactuals that can be identified with our quantum query scheme are two-way counterfactuals, since these are all the information about $\Prb(F)$ contained in $\rho_{XY}$. So the three-way joint counterfactuals here are also not identifiable with our quantum query scheme.

\section{Proving Theorem~\ref{thm:n-way}: quantum oracle advantage in partial identifiability} \label{app:partial}

\subsection{A simple example}
We first give an explicit example that illustrates Theorem~\ref{thm:n-way} in the simplest nontrivial case, namely $|\mathcal X|=3$ and $|\mathcal Y|=2$, then show how to extend it to $n := \max\{|\mathcal X|,|\mathcal Y|\}$.
This example shows that, for fixed observational and interventional data, since a quantum oracle can further identify two-way
counterfactuals, it provides a strictly tighter
upper bound on a three-way joint counterfactual than the best possible classical bound, thereby witnessing a partial-identification advantage beyond the two-way joint counterfactuals that are already identified quantumly.

Let the input take three values $X\in \{0,1,2\},$
and let the output $Y$ be binary. We parameterize the full 3-way joint counterfactual distribution
\begin{equation}
  D(y_0,y_1,y_2)
  := \Prb(Y_0=y_0,\ Y_1=y_1,\ Y_2=y_2),
\end{equation}
for $(y_0,y_1,y_2)\in\{0,1\}^3$, by
\begin{equation}
\begin{aligned}
  a &:= D(0,0,0), \quad b := D(0,0,1),  \quad c := D(0,1,0), \\
   d &:= D(0,1,1), \quad e := D(1,0,0), \quad f := D(1,0,1), \\
   g &:= D(1,1,0), \quad h := D(1,1,1).
\end{aligned}
\end{equation}
These parameters satisfy $a,b,c,d,e,f,g,h\in [0,1]$ and
\begin{equation}
  a + b + c + d + e + f + g + h = 1.
\end{equation}
The 3-way joint counterfactual we will focus on is
\begin{equation}
  h=\Prb(Y_0=1,\ Y_1=1,\ Y_2=1).
\end{equation}
We will now provide an explicit example where under the quantum constraints, $h$ has a tighter bound than that achievable under the constraints from a classical oracle.

\subsubsection{With quantum oracles:}

For concreteness, consider the case where all two-way marginals are identified to be uniform:
\begin{equation}
 \Prb(Y_i=y,\ Y_j=y')
  = \frac{1}{4}
  \quad\forall i\neq j 
  \in\{0,1,2\},\ \forall y,y'\in\{0,1\}.
  \label{eq:uniform-C}
\end{equation}
We now derive the resulting constraints on $(a,\dots,h)$ and solve them
explicitly.

The two-way marginals are obtained by:
\begin{align}
  \Prb(Y_0=y_0,Y_1=y_1)
    &= \sum_{y_2} D(y_0,y_1,y_2),\\
  \Prb(Y_0=y_0,Y_2=y_2)
    &= \sum_{y_1} D(y_0,y_1,y_2),\\
  \Prb(Y_1=y_1,Y_2=y_2)
    &= \sum_{y_0} D(y_0,y_1,y_2).
\end{align}
Imposing Eq.~\eqref{eq:uniform-C} yields the following linear
constraints.

For the pair $(Y_0,Y_1)$:
\begin{equation}
\begin{aligned}
  a + b &= \tfrac14, &\quad& (Y_0=0,Y_1=0),\\
  c + d &= \tfrac14, &\quad& (Y_0=0,Y_1=1),\\
  e + f &= \tfrac14, &\quad& (Y_0=1,Y_1=0),\\
  g + h &= \tfrac14, &\quad& (Y_0=1,Y_1=1).
\end{aligned}
\label{eq:Y0Y1-constraints}
\end{equation}

For the pair $(Y_0,Y_2)$:
\begin{equation}
\begin{aligned}
  a + c &= \tfrac14, &\quad& (Y_0=0,Y_2=0),\\
  b + d &= \tfrac14, &\quad& (Y_0=0,Y_2=1),\\
  e + g &= \tfrac14, &\quad& (Y_0=1,Y_2=0),\\
  f + h &= \tfrac14, &\quad& (Y_0=1,Y_2=1).
\end{aligned}
\label{eq:Y0Y2-constraints}
\end{equation}

For the pair $(Y_1,Y_2)$:
\begin{equation}
\begin{aligned}
  a + e &= \tfrac14, &\quad& (Y_1=0,Y_2=0),\\
  b + f &= \tfrac14, &\quad& (Y_1=0,Y_2=1),\\
  c + g &= \tfrac14, &\quad& (Y_1=1,Y_2=0),\\
  d + h &= \tfrac14, &\quad& (Y_1=1,Y_2=1).
\end{aligned}
\label{eq:Y1Y2-constraints}
\end{equation}

Solving Eqs.~\eqref{eq:Y0Y1-constraints},
\eqref{eq:Y0Y2-constraints}, \eqref{eq:Y1Y2-constraints} yields the family of distributions consistent with the quantum constraints:
\begin{equation}
  (a,b,c,d,e,f,g,h)
  = \bigl(\tfrac14-h,\ h,\ h,\ \tfrac14-h,\ h,\ \tfrac14-h,\ \tfrac14-h,\ h\bigr).
  \label{eq:quantum-family}
\end{equation}

Non-negativity of each entry amounts to
\begin{equation}
  h \ge 0,\qquad \tfrac14 - h \ge 0,
\end{equation}
so that $ 0 \le h \le \tfrac14. $

Moreover, the endpoints of this interval are achieved:

\begin{itemize}
  \item For $h=0$ we obtain
    \[
      (a,b,c,d,e,f,g,h)
      = \bigl(\tfrac14, 0, 0, \tfrac14, 0, \tfrac14, \tfrac14, 0\bigr),
    \]
    which satisfies all two-way uniformity constraints.
  \item For $h=\tfrac14$ we obtain
    \[
      (a,b,c,d,e,f,g,h)
      = \bigl(0, \tfrac14, \tfrac14, 0, \tfrac14, 0, 0, \tfrac14\bigr),
    \]
    which also satisfies all constraints.
\end{itemize}

Thus, the quantum-constrained feasible range of the 3-way joint is:
$h = \Prb(Y_0=1,Y_1=1,Y_2=1) \in [0,\tfrac14] $.

\subsubsection{With classical oracles}

We now show that the corresponding classical feasible set allows for a
strictly larger upper bound on $h$.

Classically, with only interventional data (or, in the no-confounding case, observational conditionals) one may at best constrain the
\emph{one-way} counterfactuals $\Prb(Y_x=y)$.
In this example, we assume the symmetric case
\begin{equation}\label{eq:oneway}
  \Prb(Y_x=1) = \tfrac12
  \quad\text{for }x=0,1,2,
\end{equation}
and no information about two-way marginals. 

Expressing the single-marginal constraints in terms of $(a,\dots,h)$ one has the following:
\begin{align}
  \Prb(Y_0=1) &= e+f+g+h = \tfrac12,\\
  \Prb(Y_1=1) &= c+d+g+h = \tfrac12,\\
  \Prb(Y_2=1) &= b+d+f+h = \tfrac12.
\end{align}

Consider the perfectly correlated model
\begin{equation}\label{eq:perfectco}
  D(0,0,0) = \tfrac12,\qquad D(1,1,1) = \tfrac12,
\end{equation}
with all other $D(y_0,y_1,y_2)=0$. In terms of
$(a,b,c,d,e,f,g,h)$, this is
\begin{equation}
  (a,b,c,d,e,f,g,h) = \bigl(\tfrac12, 0,0,0,0,0,0,\tfrac12\bigr).
\end{equation}
Each $Y_x$ is equal to $1$ with probability $1/2$, so the
single-marginal constraints are satisfied. But, for this model,
$ h = \tfrac12.$ Thus the quantum oracles yield a strictly tighter upper bound on this three-way joint counterfactual than is possible classically.

\subsection{Extension to general cardinality $n$}

Finally, we show how this example extends to arbitrary
cardinality $n := \max\{|\mathcal X|,|\mathcal Y|\}$. Suppose $|\mathcal X|=n\ge 3$ and $|\mathcal Y|=2$. Consider the $n$-way joint counterfactual:
\begin{equation}
  \Prb\bigl(Y_0=1,\ Y_1=1,\ Y_2=1,\ Y_3= y_3,\dots,Y_{n-1}= y_{n-1}\bigr),
\end{equation}
for some fixed $( y_3,\dots,y_{n-1})\in\{0,1\}^{n-3}$.

 Consider further the special case where $\Prb(F)$ is such that it is only supported on the cases where $f$ is a fixed constant function on $X=3,4,\dots, n-1$. Specifically, it maps $( 3,4,\dots, n-1)$ to $(y_3,\dots,y_{n-1})$ and thus,
\begin{align}
    &\Prb\bigl(Y_0=a,\ Y_1=b,\ Y_2=c,\ Y_3= y_3,\dots,Y_{n-1}=y_{n-1}\bigr) \nonumber \\
    =&\Prb\bigl(Y_0=a,\ Y_1=b,\ Y_2=c,\bigr), \,\, \forall a,b,c \in\{0,1\}
\end{align} 
Classically, with only one-way marginals fixed  (in particular, assume that they are given by \cref{eq:oneway} and $\Prb(Y_i=y_i)=1$ for $i=3,\dots,n-1$),  one can
realize models in which $(Y_0,Y_1,Y_2)$ are perfectly correlated as in \cref{eq:perfectco}, and hence the above $n$-way joint attains $\tfrac12$. Quantumly, imposing all two-way marginals as in \cref{eq:uniform-C} restricts the feasible
family to the form \eqref{eq:quantum-family}, hence upperbounding the $n$-way joint by $\tfrac14$.

Thus, for each $n := \max\{|\mathcal X|,|\mathcal Y|\}\geq2$, there exist cases where an $n$-way joint counterfactual has a strictly tighter quantum upper bound. 

\end{document}